\documentclass[aps,twocolumn]{revtex4}
\usepackage{epsfig}
\usepackage{bm}
   \begin{document}
   \def\I.#1{\it #1}
  \newcommand{\B}[1]{{\bm{#1}}}
   \def\C.#1{{\cal #1}}
\title{New Algorithm for Parallel Laplacian Growth by Iterated Conformal Maps}
   \author {Anders Levermann and Itamar Procaccia}
   \address{Department of~~Chemical Physics, The Weizmann Institute of
   Science, Rehovot 76100, Israel}

%
%
\begin{abstract}
We report a new algorithm to generate Laplacian Growth Patterns
using iterated conformal maps. The difficulty of growing
a complete layer with local width proportional to the gradient of the
Laplacian field is overcome. The resulting growth patterns are compared
to those obtained by the best algorithms of direct numerical solutions.
The fractal dimension of the patterns is discussed.
\end{abstract}
  \maketitle

Laplacian Growth Patterns are obtained when the boundary $\Gamma$ of
a 2-dimensional domain is grown
at a rate proportional to the gradient of a Laplacian field $P$ \cite{84Pat}.
The classic examples of such patterns appear in viscous fingering
in constrained geometries (like Hele-Shaw cells or porous media).
Here a less viscous fluid (inside the domain bounded by $\Gamma$)
displaces a more viscous fluid which is outside the domain.
The field $P$ is the pressure, and Darcy's law determines
the velocity $\B v$ to be proportional to $\B \nabla P$.
Using the incompressibility constraint
outside the domain
$\nabla^2 P=0$, and each point of $\Gamma$ is advanced at a rate
proportional to $\B \nabla P$ \cite{58ST,86BKT}. Thus in numerical
algorithms \cite{94HLS} one needs at each time step to add on a whole layer to the pattern
with a local width proportional to $|\B \nabla P|$.
The boundary conditions are such that
in radial geometry as $r\to \infty$ the flux
is $\B \nabla P={\rm const}\times\hat r/r$. On the boundary $\Gamma$
one usually solves the problem with
the condition $P=\sigma\kappa$ where $\sigma$
is the surface tension and $\kappa$ the local
curvature of $\Gamma$ \cite{86BKT}. Without this
(or some other) ultraviolet regularization
Laplacian Growth reaches a singularity (in the form of a cusp) in finite time \cite{84SB}.
In this Letter we present a new algorithm to grow such patterns using
iterated conformal maps. The basic method was introduced \cite{98HL} in the context of
Diffusion Limited Aggregation (DLA) \cite{81WS} where it was successfully
employed to solve a number of outstanding problems in the theory of
DLA \cite{99DHOPSS,00DFHP}, its fractal dimension \cite{00DP,00DLP} and its multifractal
properties \cite{01BJLMP,02JLMP,03JMP}.

At the heart of the method stands the elementary map
$\phi_{\lambda,\theta}$ which transforms the unit
circle to a circle with a ``bump" of linear size $\sqrt{\lambda}$
around the point $w=e^{i\theta}$. We employ the elementary map
\cite{98HL}
\begin{eqnarray}
   &&\phi_{\lambda,0}(w) = \sqrt{w} \left\{ \frac{(1+
   \lambda)}{2w}(1+w)\right.    \label{eq-f}\\
   &&\left.\times \left [ 1+w+w \left( 1+\frac{1}{w^2} -\frac{2}{w}
\frac{1-\lambda} {1+ \lambda} \right) ^{1/2} \right] -1 \right \} ^{1/2} \nonumber\\
   &&\phi_{\lambda,\theta} (w) = e^{i \theta} \phi_{\lambda,0}(e^{-i
   \theta}
   w) \ . \nonumber
\end{eqnarray}
This map grows a semi-circular bump with two branch points at
the angular positions $\theta\pm\alpha$, where
\begin{equation}
\alpha = \tan^{-1} \left( \frac{2\sqrt{\lambda}}{1-\lambda}\right) \ . \label{bp}
\end{equation}
By iterating this fundamental map with randomly chosen angles $\theta_n$ and
the bump sizes $\lambda_n$ chosen such as to obtain equal size particles on the
cluster,
\begin{eqnarray*}
\lambda_{n}=\frac{\lambda_0}
{\left|\Phi^{{(n-1)}'}\left(e^{i\theta_{n}}\right)\right|^2} \ ,
\end{eqnarray*}
one can easily grow a DLA pattern. On the other hand,
the direct application to the closely related problem of
viscous fingering remained unaccomplished due to technical difficulties that have
been now surmounted, as we report below.

We are interested in $\Phi^{(n)}(w)$ which conformally maps the exterior of the unit
circle $e^{i\theta}$ in the mathematical $w$--plane onto the complement of the Laplacian
pattern in the physical $z$--plane. 
As in previous work the map $\Phi^{(n)}(w)$ is obtained by iteration of fundamental 
maps $\phi_{\lambda_j,\theta_j} (w)$. The superscript $n$ denotes
$n$ growth steps. The gradient of the Laplacian field
$\B \nabla P(z(s))$ is
\begin{equation}
\left| {\B \nabla P(z(s))} \right| =\frac{1}{\left|{\Phi^{(n)}}^\prime(e^{i\theta})\right|}
\ ,\quad z(s)=\Phi^{(n)}(e^{i\theta}) \ .
\label{gradient}
\end{equation}
Here $s$ is an arc-length parametrization of the boundary. Contrary
to DLA which is grown
serial, i.e. particle by particle, in Laplacian growth
for the problem of viscous fingering
we need to grow in parallel, i.e. layer by layer.
Nevertheless the map $\Phi^{(n)}(w)$ is still constructed recursively. Suppose that
we completed the last layer at growth step m, i.e. $\Phi^{(m)}(w)$ is known,
and we want to find the map $\Phi^{(m+p)}(w)$ which maps the exterior of the unit
circle to the exterior of the pattern after the addition of one more
layer whose local width is proportional to $|\B \nabla P(z(s))|$.
The number $p$ of growth events should be arranged precisely such as
to add the aforementioned layer.
To grow a full layer of non-overlapping bumps using the 
elementary map (\ref{eq-f}) is difficult
because the series of sizes $\{\lambda_{m+k}\}_{k=1}^p$ and 
of positions $\{\theta_{m+k}\}_{k=1}^p$
depend on each other.

Consider then the ($m+k$)th step of growth that we implement at the
angle $\theta_{m+k}$. Due to the re-paramterization during the 
$p$ growth steps $k=1,2,\dots p$
we need first to find an angle $\tilde\theta_{m+k}$ according to the following rule. For a
given position, the size has to be such that its image 
under the map $\Phi^{(m+k-1)}$ is proportional to the local field
$\left| \Phi^{{(m)}'}\left( e^{i\tilde\theta_{m+k}} \right)\right|^{-1}$, i.e.
\begin{eqnarray*}
\lambda_{m+k}=\frac{\lambda_0}
{\left|\Phi^{{(m+k-1)}'}\left(e^{i\theta_{m+k}}\right)\right|^2}
\cdot \left|\Phi^\prime_{m}\left( e^{i\tilde\theta_{m+k}} \right)\right|^{-2} \ .
\end{eqnarray*}
here $\tilde \theta_{m+k}$ is defined through
\begin{equation}
\Phi^{(m)}\left(e^{i\tilde\theta_{m+k}}\right)
=\Phi^{(m+k-1)}\left(e^{i\theta_{m+k}}\right) \ .
\end{equation}

On the other hand, the position has to be such that the new bump on the cluster
precisely touches the previously grown one, i.e. the image of one of its branch
points has to be equal to the image of the corresponding branch point of the previous
bump. When growing the layer in a mathematically positive direction this means that
\begin{equation} \label{the-equ-1}
\Phi^{(m+k)} \!\!\left(\!\!e^{i\left(\theta_{m+k}-\alpha_{m+k}\right)}\!\!\right) =
\Phi^{(m+k-1)} \!\!\left(\!\!e^{i\left(\theta_{m+k-1}+\alpha_{m+k-1}\right)}\!\!\right)\ ,
\end{equation}
where according to Eq.(\ref{bp})
$\alpha_n = \tan^{-1}\left( 2\sqrt{\lambda_n}/(1-\lambda_n)\right)$
Note that $\alpha_n$ only depends on the
size $\lambda_n$ and not on the position, because it is a property of the un-rotated
fundamental map, namely its branch points.
By use of the inverse fundamental map equation (\ref{the-equ-1}) simplifies to
\begin{eqnarray}
&&e^{i\theta_{m+k}} \cdot \phi_{\lambda_{m+k},0}
\left(e^{-i\theta_{m+k}}\cdot e^{i\left(\theta_{m+k}-\alpha_{m+k}\right)}\right) \nonumber\\&&=
e^{i\left(\theta_{m+k-1}+\alpha_{m+k-1}\right)} \ ,
\end{eqnarray}
where $\phi_{\lambda_{m+k},0}$ is the fundamental map which depends only on the size of
the bump $\sqrt{\lambda_{m+k}}$. This can be further simplified to get an equation for
$\theta_{m+k}$
\begin{equation} \label{the-equ-2}
e^{i\theta_{m+k}} =
\frac{e^{i\left(\theta_{m+k-1}-\alpha_{m+k-1}\right)}}
{\phi_{\lambda_{m+k},0}\left(e^{i\alpha_{m+k}}\right) } \ ,
\end{equation}
where the right hand side depends only on the previous bump $(m+k-1)$ and the
size of the new particle $(m+k)$. Thus we can compute $\theta_{m+k}$ from
equation (\ref{the-equ-2}) once we know the size of the bump.

To accomplish the goal of growing a full layer we now need to estimate
the size of the bump
$\lambda_{m+k}$ and from there compute the position $\theta_{m+k}$ according to equation
(\ref{the-equ-2}). This will exactly close the gap between successive bumps in the layer.
In estimating  $\lambda_{m+k}$ we use the fact
that the bumps are minute on the scale of the growth pattern. Therefore
the field does not vary significantly from bump to bump. Thus we set the
value of $\lambda_{m+k}$ to
\begin{equation} \label{lam_approx}
\lambda_{m+k} \equiv
\frac{\lambda_0}{\left|\Phi^{(m+k-1)~'}\left(e^{i\theta_{m+k-1}}\right)\right|^2}
\cdot \left| \Phi^\prime_{m}\left( e^{i\tilde\theta_{m+k-1}} \right)\right|^{-2} \ .
\end{equation}
In other words, we estimate $\lambda_{m+k}$ for the unknown position
$\theta_{m+k}$ at the position of the last bump $\theta_{m+k-1}$.
Note that with equation (\ref{the-equ-2}) we are guaranteed to have
a full cover of the layer independently of the estimate of $\lambda_{m+k}$.
Only the very last bump may not ``fit" in, and is therefore not grown.

In order to avoid a bias in the growth pattern we alternate the direction of filling
the layer from clockwise to anti-clockwise and vice versa. Furthermore we choose
the position of the first bump of a given layer randomly on the unit circle. Thirdly, 
to avoid growing very many very small particles in the fjords, we introduce a cut-off
$\lambda_{\rm cut}$ for $\lambda_n$. That means that if a bump is about to be
grown with a smaller $\lambda_n$ then we just avoid growing it and proceed to a
position which is $\lambda_{\rm cut}$ further down the unit circle. The results of
our algorithm were checked to be independent of $\lambda_{\rm cut}$
in the range  $10^{-6}\le \lambda_{\rm cut}\le 10^{-10}$. The results
shown in this Letter are for the lowest cut-off $\lambda_{cut}=10^{-10}$.

\begin{figure}
\centering
\includegraphics[width=0.50\textwidth]{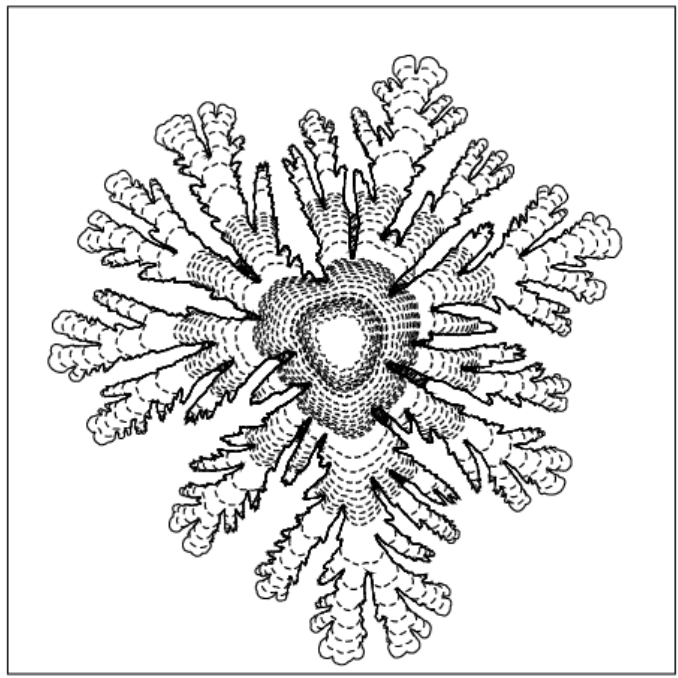}
\caption{Viscous fingering pattern obtained through
100 000 growth events. The intermediate steps of growth are also shown. Every drawn
intermediate pattern is an actual conformal map of the unit circle,
with the last one being $\Phi^{(100~000)}(e^{i\theta})$.}
\label{pattern}
\end{figure}

\begin{figure}
\centering
\includegraphics[width=0.50\textwidth]{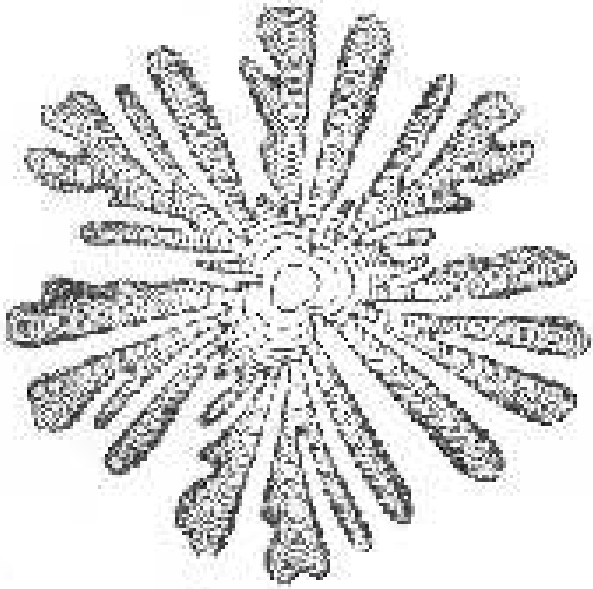}
\caption{Viscous fingering pattern obtained by direct numerical
solution, see \cite{94HLS}.}
\label{pattern2}
\end{figure}

Finally, the conformal map at the end of the growth of the
layer can be written as
\begin{equation}
\Phi^{(m+p)}(\omega) =\Phi^{(m)}\circ\phi_{\lambda_{m+1}\theta_{m+1}}\circ\dots\circ
\phi_{\lambda_{m+p}\theta_{m+p}} (\omega) \ ,
\end{equation}
where $\circ$ stands for a functional composition. In Fig.
\ref{pattern} we present $\Phi^{(100~000)}(\omega)$ which is a growth
pattern with 100 000 growth events, including the intermediate stages
of growth. The growth patterns is very
similar to those obtained by direct numerical simulations of
viscous fingering in a radial geometry. For comparison we show in
Fig. \ref{pattern2} the pattern obtained by direct numerical solution
\cite{94HLS}.

\begin{figure}
\centering
\includegraphics[width=0.50\textwidth]{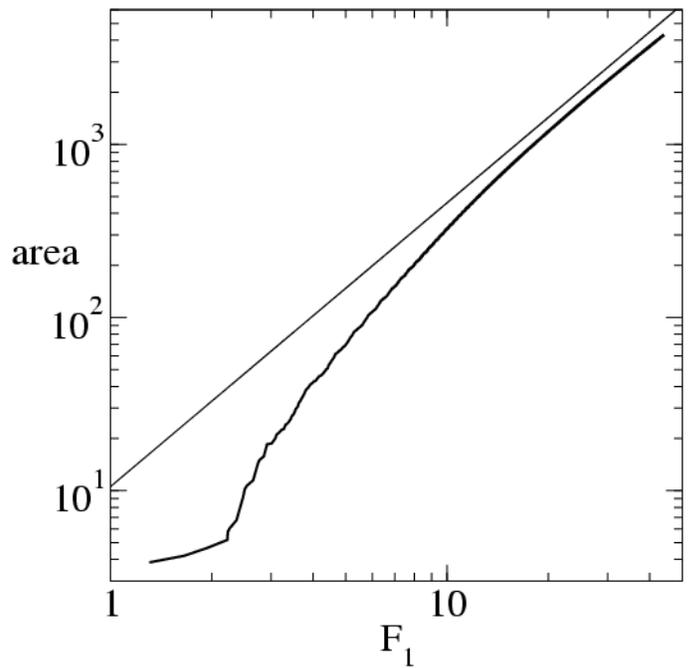}
\caption{Double logarithmic plot of the area $S^{(n)} $ vs. $F_1^{(n)}$}
\label{arearad}
\end{figure}

One advantage of the present approach is that the conformal map $\Phi^{(n)}(\omega)$
is given explicitly in terms of an iteration of analytically known fundamental maps,
providing us with an analytic control on the grown clusters. The map
$\Phi^{(n)}(\omega)$ admits a Laurent expansion
\begin{equation}
   \Phi^{(n)}(\omega) = F_1^{(n)} \omega +F_0^{(n)}
   +\frac{F_{-1}^{(n)}}{\omega}+~\cdots \ .
\end{equation}
The coefficient of the linear term is the Laplace radius. Defining the
radius of the minimal circle that contains the growth pattern by
$R_n$, one has the rigorous result that
\begin{equation}
R_n\ge F_1^{(n)} \ge R_n/4 \ .
\end{equation}
It is therefore natural to define the fractal dimension of the
cluster by the radius-area scaling relation
\begin{equation}
   F_1^{(n)}\sim S^{1/D} \ ,
\end{equation}
where $S^{(n)}$ is the area of the cluster,
\begin{equation}
S^{(n)}= \sum_{j=1}^n \lambda_j~ |{\Phi^{(j-1)}}' (e^{i\theta_j})|^2 \ .
\end{equation}
On the other hand $F_1^{(n)}$ is given analytically by
\begin{equation}
   F_1^{(n)} = \prod_{k=1}^n \sqrt{(1+\lambda_k)} \ ,
\end{equation}
and therefore can be determined very accurately. In Fig. \ref{arearad}
we plot the area $S^{(n)}$ in double logarithmic plot against $F_1^{(n)}$.
The local slope is the apparent dimension at that cluster size. it appears that
the dimension asymptotes to the value indicated by the straight line which
is $D=1.7$.

A few comments are in order. First, our algorithm does not employ
surface tension; rather, we have a typical length scale $\lambda_0$
that removes the putative singularities. 
The boundary conditions for the Laplacian field are still zero on the cluster.
The apparent correspondence of our
patterns with those grown with surface tension regularization points out
in favor of universality with respect to the method of ultra-violet regularization.
Second, one may worry that growing semi-circular bumps means that the elementary
step has both length and height proportional to $|\B \nabla P|$, and not just
height. Since our elementary growth events are so small in spatial extent this
does not appear to be a real worry. Lastly, in previous work \cite{01BDLP,02BDP,02HLP} the
technical difficulty of growing a complete layer was circumvented by
constructing a family of models which included Laplacian growth only
as a limiting model. This family of models achieved a partial coverage
${\cal C}\le 0.65$ of every layer of growth, with
${\cal C}=1$ being Parallel Laplacian growth which could be considered only as
an extrapolation of lower values of ${\cal C}$. Assuming monotonicity
of geometric properties as a function of ${\cal C}$,
the extrapolation procedure indicated strongly that viscous fingering were not in
the same universality class as DLA, having an extrapolated dimension
$D=2$. The present algorithm which achieves directly the limit ${\cal C}=1$
appears at odds with the extrapolation procedure. The direct measurement
of the dimension is in close correspondence with DLA whose dimension is
$1.713\pm 0.03$ \cite{00DLP}. At present it is not clear whether the contradiction
is due to a non-monotonicity as a function of ${\cal C}$,
Whether the clusters have not reached their asymptotic properties
or whether there is another reason that will be illuminated by further research.

At any rate it is hoped that the availability of a direct procedure to
grow viscous fingering patterns by iterated conformal maps will help
to achieve a theoretical understanding of this problem on the same level
of the understanding of DLA.

\end{document}